# Quantum Hall Effect in Ultrahigh Mobility Two-dimensional Hole Gas of Black Phosphorus

**Authors:** *Gen Long[1,†], Denis Maryenko[2,†], Junying Shen[1,†], Shuigang Xu[1], Jianqiang Hou[1], Zefei Wu[1], Wingki Wong[1], Tianyi Han[1], Jiangxiazi Lin[1], Yuan Cai[1], Rolf Lortz[1], Ning Wang[1,\*]*

**Affiliations:**

*1. Department of Physics and Center for 1D/2D Quantum Materials , The Hong Kong University of Science and Technology, Clear Water Bay, Hong Kong, China*

*2. RIKEN Center for Emergent Matter Science (CEMS), Wako 351-0198, Japan*

**\***Correspondence to: phwang@ust.hk

† These authors contributed equally to this work.

**Abstract:**

**We demonstrate that a field effect transistor (FET) made of few layer black phosphorus (BP) encapsulated in hexagonal boron nitride (h-BN) in vacuum, exhibts the room temperature hole mobility of 5200 $cm^2$/Vs being limited just by the phonon scattering. At cryogenic tempeature the FET mobility increases up to 45,000 $cm^2$/Vs, which is eight times higher compared with the mobility obtained in earlier reports. The unprecedentedly clean h-BN/BP/h-BN heterostructure exhibits Shubnikov-de Haas oscillations and quantum Hall effect with Landau level (LL) filling factors down to v=2 in conventional laboratory magnetic fields. Moreover, carrier density independent effective mass m=0.26 $m_0$ is**

**measured, and Landé g-factor g=2.47 is reported. Furthermore, an indication for a distinct hole transport behavior with up and down spin orientation is found.**

**Main Text:**

Few-layer black phosphorus (BP) has received in recent years much attention due to its unique properties making this layered material attractive for technological applications(*1-3*). This two-dimensional crystal has an anisotropic structure (Fig.1a) and is characterized by a BP thickness dependent direct band gap(*4*). In contrast to graphene, the presence of a band gap in BP permits for a selective depletion of charge carriers by electrostatic gating, which is an essential feature in field effect transistors (FETs). A high charge carrier mobility reaching 1000 cm$^2$/Vs at room temperature accentuates this material for applications at room temperature(*5*). However, the exposure of BP crystals to ambient conditions causes the oxidation of BP and significantly degrades the quality of BP channels. Nevertheless, the encapsulation of BP layers by hexagonal boron nitride (h-BN) sheets in air or in an inert gas environment is found to be very effective for preventing BP oxidation(*6-8*). Surface impurity effects are largely reduced, and high charge carrier mobility up to several 10$^3$ cm$^2$/Vs has been obtained in BP FETs at cryogenic temperature(*6-8*). The charge carrier scattering at the impurities encapsulated along with the BP layers hinders further mobility increase. Figure 1a shows the scheme of the FET fabricated from the h-BN/BP/h-BN heterostructure. Different from other fabrication methods widely reported in the literatures, our h-BN/BP/h-BN heterostructure is ensembled by encapsulating BP crystal between h-BN sheets in vacuum conditions ($P$=10$^{-3}$ Torr). Such a fabrication method reduces the surface absorption from the environment and thus minimizes the impurities formed at the BP interfaces. A device fabricated under such conditions shows also distinct transport characteristics.

Figure 1b exemplifies the dependence of the FET conductance at T=2 K on the back gate voltage $V_g$ applied between the Si substrate and the source contact. The gradual conductance increases with the lowering $V_g$ is accompanied with the increase of the hole carrier density $p$ shown on top panel. Field effect mobility $\mu_{FET}$ increases with the negative gate voltage. In the linear regime of the FET operation (plateau of $\mu_{FET}$ in Fig. 1b) $\mu_{FET}$ reaches 45,000 cm$^2$V$^{-1}$s$^{-1}$. The temperature dependence of $\mu_{FET}$ and the Hall mobility $\mu_H$ evaluated from the sample conductance at zero magnetic field are compared in Fig. 1c. $\mu_H$ and $\mu_{FET}$ show qualitatively the same temperature characteristics. At low temperature, $\mu_H$ reaches 25,000 cm$^2$V$^{-1}$s$^{-1}$ and shows a weak dependence on the carrier density $p$ (inset Fig.1c). The mobility values are more than four times larger compared with that in previous studies, which indicates the improved quality of h-BN/BP interfaces(7). In spite of using the advanced fabrication technique, $\mu_{FET}$ and $\mu_H$ saturate at T<20 K, which implies that the disorder scattering dominates over the phonon scattering in this temperature regime, which limits the hole mobility at cryogenic temperature(9). The increase of $\mu_H$ with the carrier density $p$ (Fig. 1c) suggests that the disorder potential is likely created by residual impurities and can be screened by the mobile carriers(7, 10, 11). The scattering behavior changes at high temperatures (T>100 K). $\mu_{FET}$ and $\mu_H$ decrease with increasing $T$ and follow the dependence $\mu \sim T^{-\gamma}$, where $\gamma$ =1.9 and 2.0 characterize the dependence for $\mu_H$ and $\mu_{FET}$, respectively (black line in Fig. 1c). The large $\gamma$ values imply that the acoustic phonon rather than the optical phonon scattering dominates over the scattering by the residual impurities in this temperature regime. It is very noticeable that the room temperature hole mobility $\mu_H$ = 5200 cm$^2$/Vs closely approaches the theoretically predicted hole mobility for a clean five-layer BP sheets, which lies in the range between 4,800 cm$^2$V$^{-1}$s$^{-1}$ and 6,400 cm$^2$V$^{-1}$s$^{-1}$(9). The realization of the predicated mobility value, which is solely limited by the phonon scattering at room temperature, is another demonstration of the improved BP heterostructure quality.

**Quantum Hall Effect (QHE) in BP 2DHG**

Figure 2a shows Hall resistance $R_{yx}$ and magnetoresistance $R_{xx}$ as a function of the magnetic field, which is measured in a clean heterostructure at the base temperature of the experiment ($T=2$ K) and at the gate voltage Vg= −60 V corresponding to the highest charge carrier mobility. $R_{yx}$ exhibits a clear plateau structure that coincides with the minima in $R_{xx}$ oscillations. $R_{yx}$ at the plateau assumes the values h/e²ν, where ν is an integer number. Thus, Landau level (LL) filling factor ν can be unambiguosly assigned to each plateau. Sweeping the back gate voltage $V_g$ at a fixed magnetic field, as shown in Fig. 2b for $B=14$T, allows the reduction of the charge carrier density gradually and reveals $R_{yx}$ plateau formation at both even and odd integer filling factors from ν =12 to ν=2. QHE observations at even and odd integer ν indicate the lifting of twofold spin degeneracy at high magnetic field. $R_{yx}$ at ν=2 plateau deviates from the exact quantization value, which is solely caused by the degradation of the ohmic contact quality at low carrier concentrations. $R_{xx}$ minima at all integer ν do not reach a zero value because of the thermal activated transport resulting from the high base temperature of our experiment (Supplemantary Materials). Inspit of this the presented transport characteristics manifest QHE in the state-of-the-art BP-heterostructure, and the thermal activtion energies (energy gaps between LL) for ν=14 at $B=10.47$T and ν=12 at $B=12.13$ T are determined to be 0.71 and 0.97 meV respectively according to the thermal activated $R_{xx}$ minima(Supplementary Materials).

Now we draw our attentention to the transport at low magnetic fields. The oscillating behavior in $R_{xx}$ (Fig. 2a) sets on at $B=4$ T, which exhibits the sequence of only even filling factors. The splitting of $R_{xx}$ oscillations at higher field occurs and develops rapidly in the magnetic field, suggesting that the exchange interaction strongly affects the splitting(*12-14*). Such a behavior signals the lifting of the spin degeneracy and adds the $R_{yx}$ quantization at odd-filling factors. The Fourier transformation of $R_{xx}$ oscillations (Fig. 2c) reveals two oscillation frequencies confirming both the lifting of spin degeneracy and the absence of another high mobility parallel conducting channel. Thus, such a system lends itself to probe the mass and the Lande g-factor of holes in BP layers.

**Effective mass, spin susceptibilty, and Lande g-factor in BP 2DHG**

The most commonly reported values of effective hole mass in BP are scattered around the value 0.26 $m_0$, where $m_0$ is the electron free mass (Table S1)(*6, 8, 15*). Analyzing the temperature dependence of SdH oscilllation amplitude in several our BP devices at varying carrier densities and along two crystal directions (Supplemantary Materials), we evaluate the hole effective mass 0.26 $m_0$ (Table 1), which is thus in agreement with previous reports.

This study measured for the first time the Landé g-factor by employing the convertional method, in which the spin resolved LLs are brought to the coincidence by tilting the sample in the magnetic field(*16-20*). This method takes advantage of the fact that the Zeeman energy $E_z = g\mu_B B_{total}$ depends on the total magnetc field $B_{total}$, whereas the cyclotron energy $E_c = \hbar e\mu_B B_\perp/m^*$ is given by the field component $B_\perp$ perpendicular to the 2DHG plane. When the Zeeman energy is a multiple integer of cyclotron energy, i.e. $E_z=iE_c$, the spin resolved LLs will overlap (energy gaps at even/odd filling factors close). A coincidence angle $\theta_c$ offers a relation for evaluating spin-susceptibility $\chi_s = gm^* = 2i\cos(\theta_c)$, where *i* is the LL coincidence index. Fig. 3b shows the evolution of carrier energy levels under a constant $B_\perp$ as $B_{total}$ and/or tilt angle increases. The dashed blue line in Fig. 3b indicates the LL coincidence event. An energy gap closing in the experiment is marked by the vanishing of SdH oscillation components of quantum states with even/odd filling factors. The FET device $R_{xx}$ is recorded at 1.6 K under a magnetic field $B_{total}$ =14T at various tilt angles $\theta$, as shown in Fig.3a. When the tilt angle approaches 71.3° (marked by the blue dashed rectangle in Fig.3a), the observed even-v $R_{xx}$ dips become weak and finally change into peaks, and those for odd filling factors align at the reference dashed lines, which demonstrate that the cyclotron component of quantum states with even filling factors vanish and signals the first LL coincidence (i.e., $i$=1). Thus, the spin-susceptibility is evaluated, $\chi_s = gm^* = 2\cos(\theta_c) = 0.641$. This study adapts the average value of the effective mass m*=0.26 $m_0$ and obtain g=2.47, which is inconsistent with the theoretically obtained effective g-factor and larger than a bare hole *g*-factor of 2(*7*). The g-factor enhancement can be caused by the exchange interaction between spins with unalike orientations(*16, 21, 22*). The exchange interaction that induced enhancements of Lande *g*-factor (spin susceptibility) have been observed in other 2DEG/2DHG systems(*23-26*). Previous studies have demonstrated an isotropic Lande g-factor in BP that is due to the weak spin-orbit coupling strengths(*7*).

**Spin-selective quantum scattering in BP 2DHG**

Figure 3a depicts different tilt angle dependences of oscillation amplitudes ($\Delta R_{xx}$) of quantum states with even and odd filling factors. For tilt angle <45°, $\Delta R_{xx}$ at even filling factors decreases dramatically with the increasing tilt angles, whereas $\Delta R_{xx}$ at odd filling factors remains unchanged. To understand the different behaviors of carriers with different spin orientations when Zeeman splitting becomes resoluble, one has to unravel individual components of carriers with different spin orientations from observed SdH oscillations. On the phenomenological level and when the interaction effects are neglected, the SdH oscillations can be described using Lifshitz-Kosevitch (LK) formalism, as follows:

$$\Delta R_{xx} = 2R_0 \sum_{r,\uparrow,\downarrow} \frac{r\lambda(T)}{\sinh(r\lambda(T))} \exp(-r\frac{\pi}{\omega\tau_{\uparrow,\downarrow}}) \cos(r\phi_{\uparrow,\downarrow}) \quad \ldots \ldots \ldots \quad (1)$$

where $\omega = eB/m^*$ is the spin-independent hole cyclotron frequency, and $\lambda(T) = 2\pi^2 k_B T / \hbar\omega$ is a function of temperature $T$. The phase $\phi_{\uparrow,\downarrow} = 2\pi(B_F/B)\{1 \mp [(g\mu_B B)/(2\hbar\omega B_F/B)]\} - \pi$ contains the term of Zeeman component $g\mu_B B$ (27, 28). $B_F$ represents the periods of SdH oscillations extracted from the FFT analysis. $\tau_{\uparrow,\downarrow}$ are the scattering times for two spin orientations.

Red solid lines in Figs. 4a and 4b display the oscillatory part of experimental $R_{xx}$ at a zero tilt angle. The blue dashed lines are the best descriptions of the experimental $\Delta R_{xx}$ using the LK formalism presented by Eq.1 considering $r = 1\ldots20$. The spin-resolved LK model reproduces the alternative oscillation amplitudes and captures the different scattering times for ↑ and ↓ spin orientations. The $\tau_\uparrow$ for spin up carriers, which are aligned with the spin of the lowest energy Lande level and have low Zeeman energy, are larger than $\tau_\downarrow$ for spin down carriers. The differences of oscillation amplitudes can be interpreted as distinct scattering rates for two spin orientations(27). The scattering times $\tau_\uparrow$ and $\tau_\downarrow$ have weak dependence on the gate voltage shown in Fig. 4c. The difference between $\tau_\uparrow$ and $\tau_\downarrow$ is also seen in the temperature dependence shown in Fig. 4d. Thus, the charge carriers in BP 2DHG show spin selective scattering behavior on a phenomenological level. This spin-selectivity in GaAs has been proposed to be caused by spin-dependent coupling between the edge states and the bulk states(29). The indication of such coupling was seen in the asymmetric response of SdH amplitude of ↑ and ↓ spins to the voltage applied between the source and the drain contacts. This behavior is absent in the BP FET (Supplementary Materials). Therefore, there can be another mechanism, which causes such a spin dependent transport behavior. In ZnO it was proposed that the exchange interaction can lead to the spin dependent scattering accompanied by the formation of non-trivial spin textures of high mobile electrons(*27*). The similar scattering mechanism proposed in ZnO can be active in the BP FETs because the parameters of ZnO and BP are comparable, and their energy scales are comparable.

This study has achieved a record high carrier mobility in a few-layer BP FET that is fabricated based on h-BN encapsulation under vacuum conditions. High-quality BP FETs show clear QHE in laboratory magnetic field. The LL crossing has been studied using the standard coincidence technique, and some quantum transport related parameters, such as the effective mass, spin-susceptibility, and Lande g-factor have been determined accordingly. A spin-selective quantum scattering mechanism is proposed to interpret the unexpected experimental data.

**Acknowledgement:**

Financial support from the Research Grants Council of Hong Kong (Project Nos. 16302215, HKU9/CRF/13G, 604112 and N_HKUST613/12) and technical support of the Raith-HKUST Nanotechnology Laboratory for the electron-beam lithography facility at MCPF are hereby acknowledged.

**Author contributions:** N. Wang and G. Long conceived the project. G. Long fabricated the devices and performed cryogenic measurements with the help of J. Y. Shen and S. G. Xu. G. Long, D. Maryenko and N. Wang analyzed the data, and wrote the manuscript. Other authors provided technical assistance in the project.

**Competing financial interests:**

The authors declare no competing financial interests.

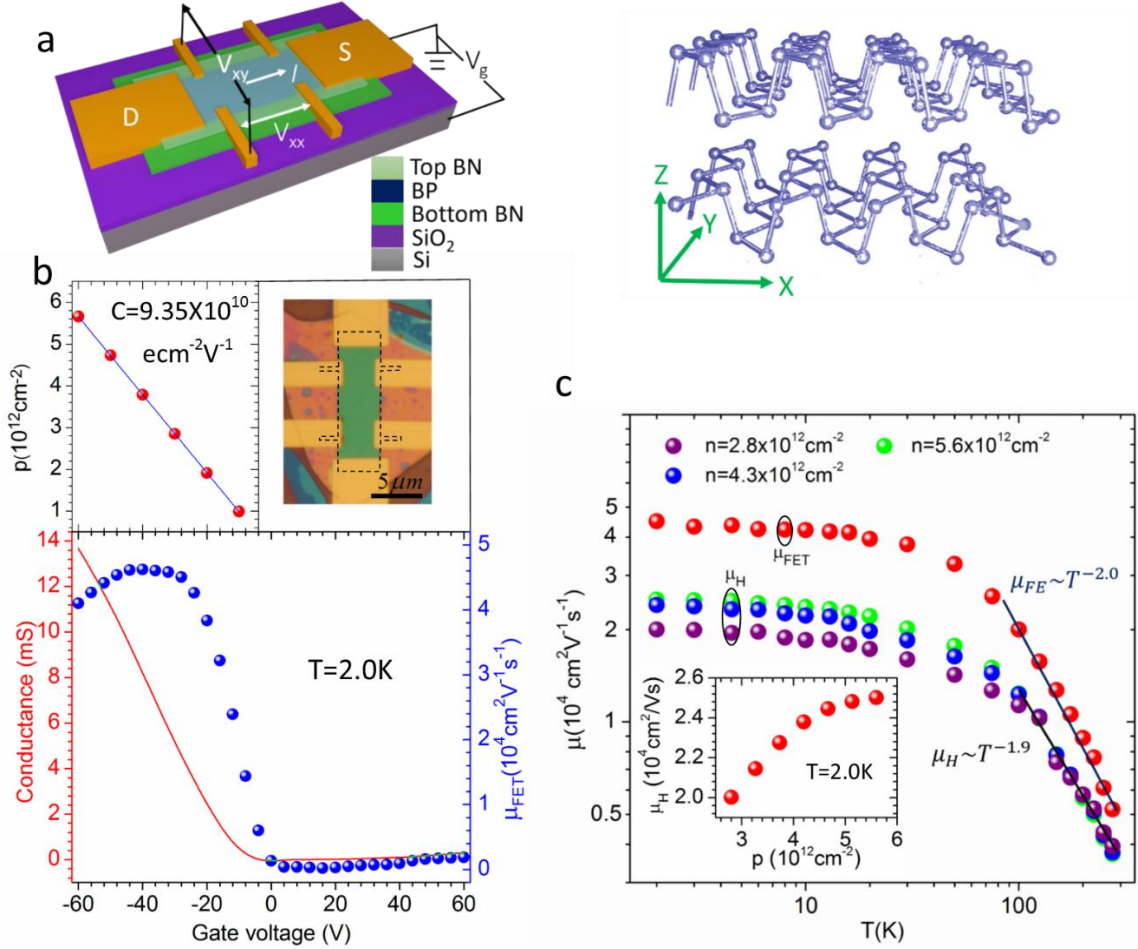

**Fig.1 Device characteristics of black phosphorus field-effect devices.** (**a**) Left panel shows the schematic view of the BP FET devices and the measurement configuration. The right panel shows the atomic structure of BP. (**b**) Cryogenic temperature (2K) conductance-gate voltage characteristics ($V_{ds}$=1mV, red line) and FET mobility at varying gate voltages (blue dots). FET mobility is given by $\mu_{FET} = \frac{d\sigma}{dV_g}\frac{L}{W}\frac{1}{C}$, where L and W denote the length and the width of devices, respectively, C denotes the capacitance deduced from upper left panel. The upper left panel shows carrier densities obtained from the oscillation periods. The blue line represents the linear fit of carrier density change with the gate voltage. The slope of the blue line corresponds to a capacitance C= $9.35*10^{10}$ ecm$^{-2}$V$^{-1}$. The upper right panel is an optical micrograph of a typical BP FETs. The scale bar is 5 um. (**c**) FE mobility (red) and Hall mobility at different carrier densities (green: 5.6; blue: 4.3; purple: $2.8 \times 10^{12}$cm$^{-2}$) as a functions of temperature. Linear

fitting were applied to extract $\gamma$ in the phonon limited region (100K~275K). The inset shows the dependence of Hall mobility on carrier density. Hall mobility is defined as $\mu_{Hall} = \dfrac{\sigma}{p \cdot e} \dfrac{L}{W}$, where the carrier density $p$ shown in panel c is estimated from the Hall effect.

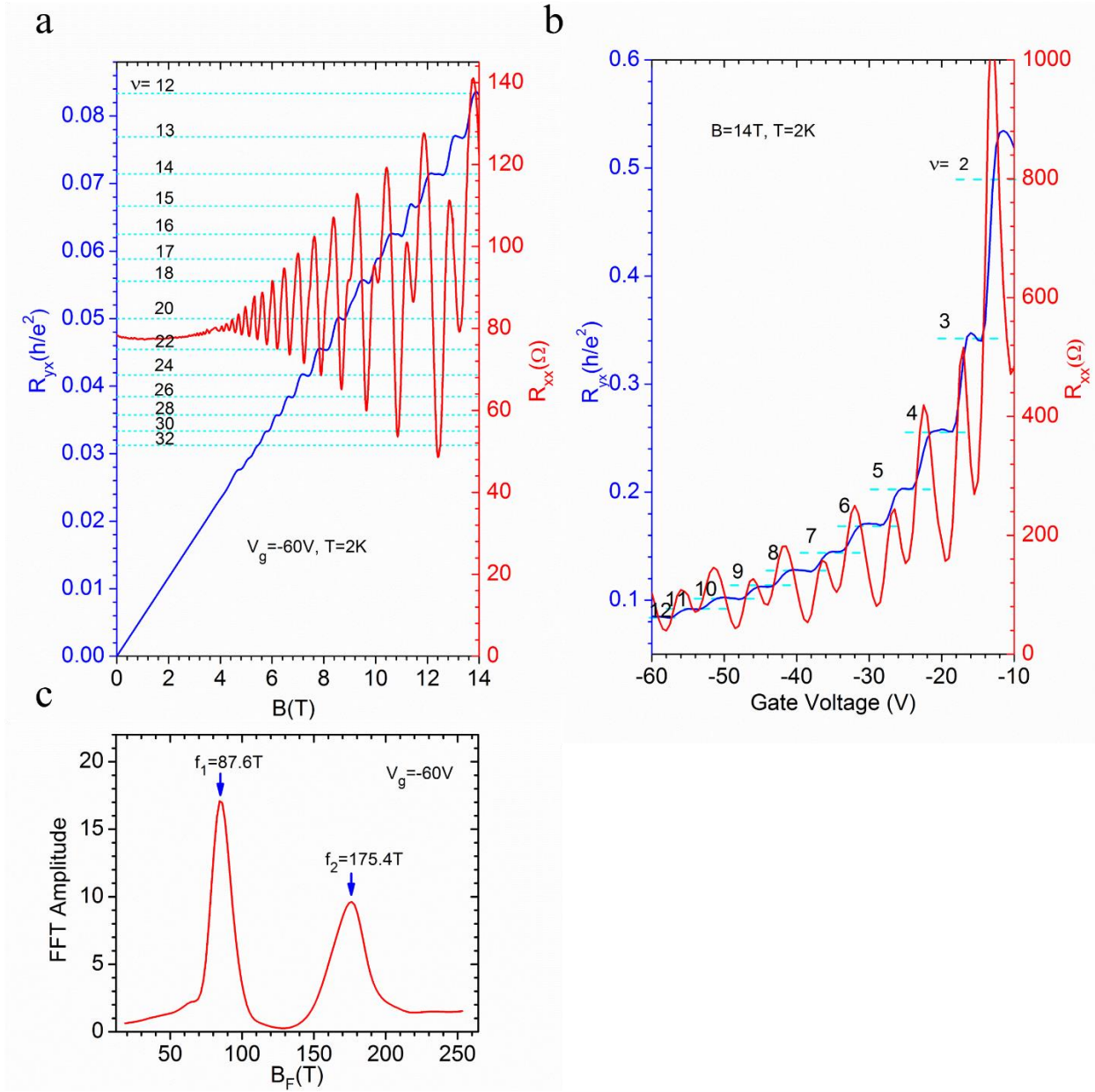

**Fig.2 Quantum Hall effect in black phosphorus 2DHG.** **(a)** Hall resistance Ryx (red) and magnetoresistance Rxx (blue) as a function of magnetic field at the gate voltage Vg= -60V and temperature T=2K. **(b)** Ryx (blue) and Rxx (red) as a function of gate voltages at T=2K under a magnetic field of 14T. Horizontal dashed lines in (a) and (b) mark the integer filling factors $v$ at quantized Ryx=h/ve$^2$. Quantum states with filling factors from 2 to 12 are observed. **(c)** FFT analysis of magnetoresistance oscillations measured at a gate voltage -60V.

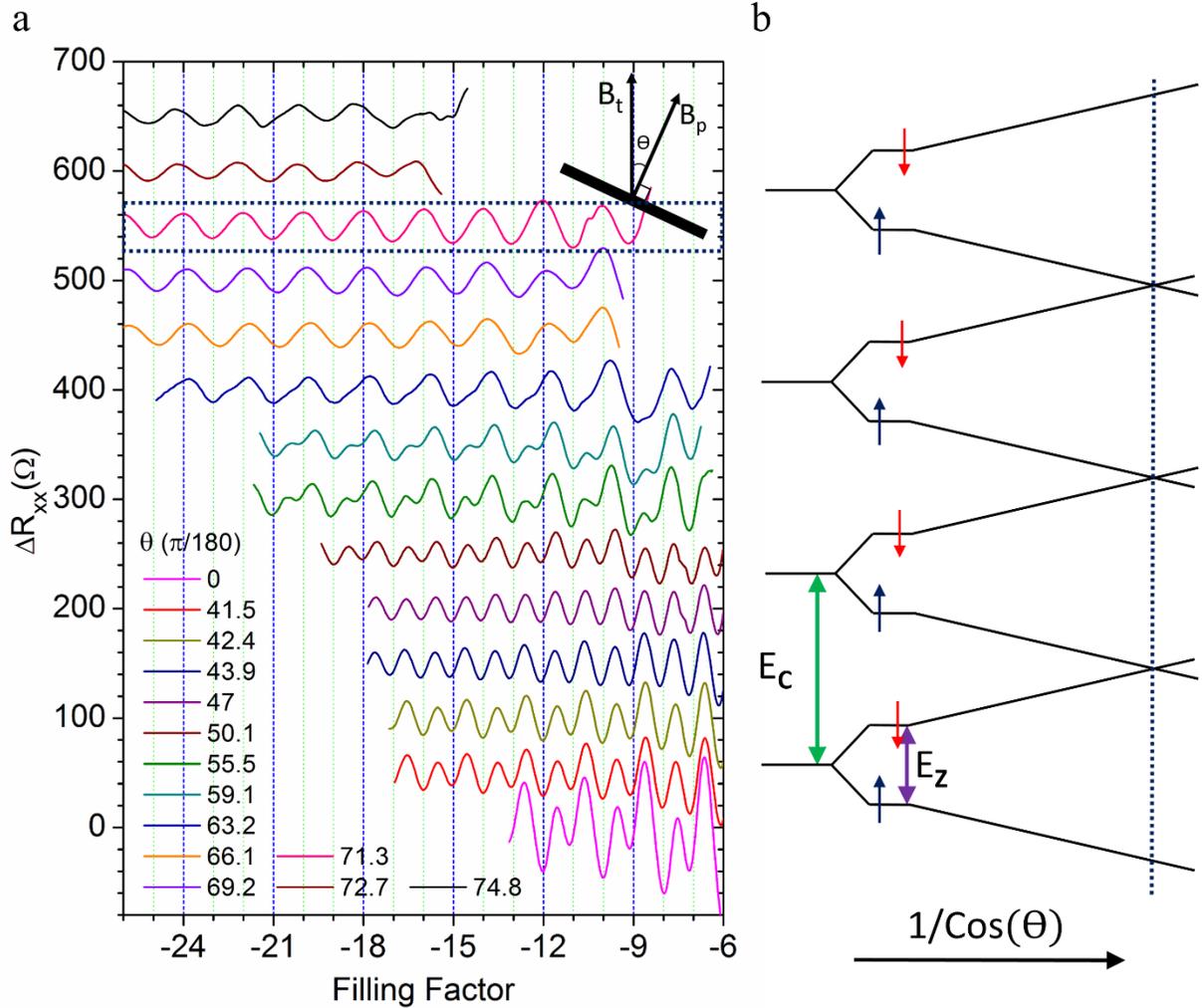

**Fig.3 Probing Landau level crossing through the coincidence technique. (a)** Magnetoresistance plotted as a function of filling factors showing the evolution of SdH oscillations with tilt angles under a magnetic field of 14T. Vertical dashed lines mark the filling factors from 7 to 26. Blue dashed rectangle marks the SdH oscillation when the cyclotron energy

gap closes. The top right inset shows the measurement configuration. **(b)** Schematic fan diagram showing Zeeman splitting Landau levels at different tilt angles. The vertical blue dashed line shows the coincidence angle for the cyclotron energy gap closing. The green solid line represents the cyclotron energy $E_c = \hbar\omega$; the purple solid line represent the Zeeman splitting energy $E_z = g\mu_B B$. The blue upward arrows (spin up) and red downward arrows (spin down) show the spin properties of the carriers occupying the corresponding energy levels.

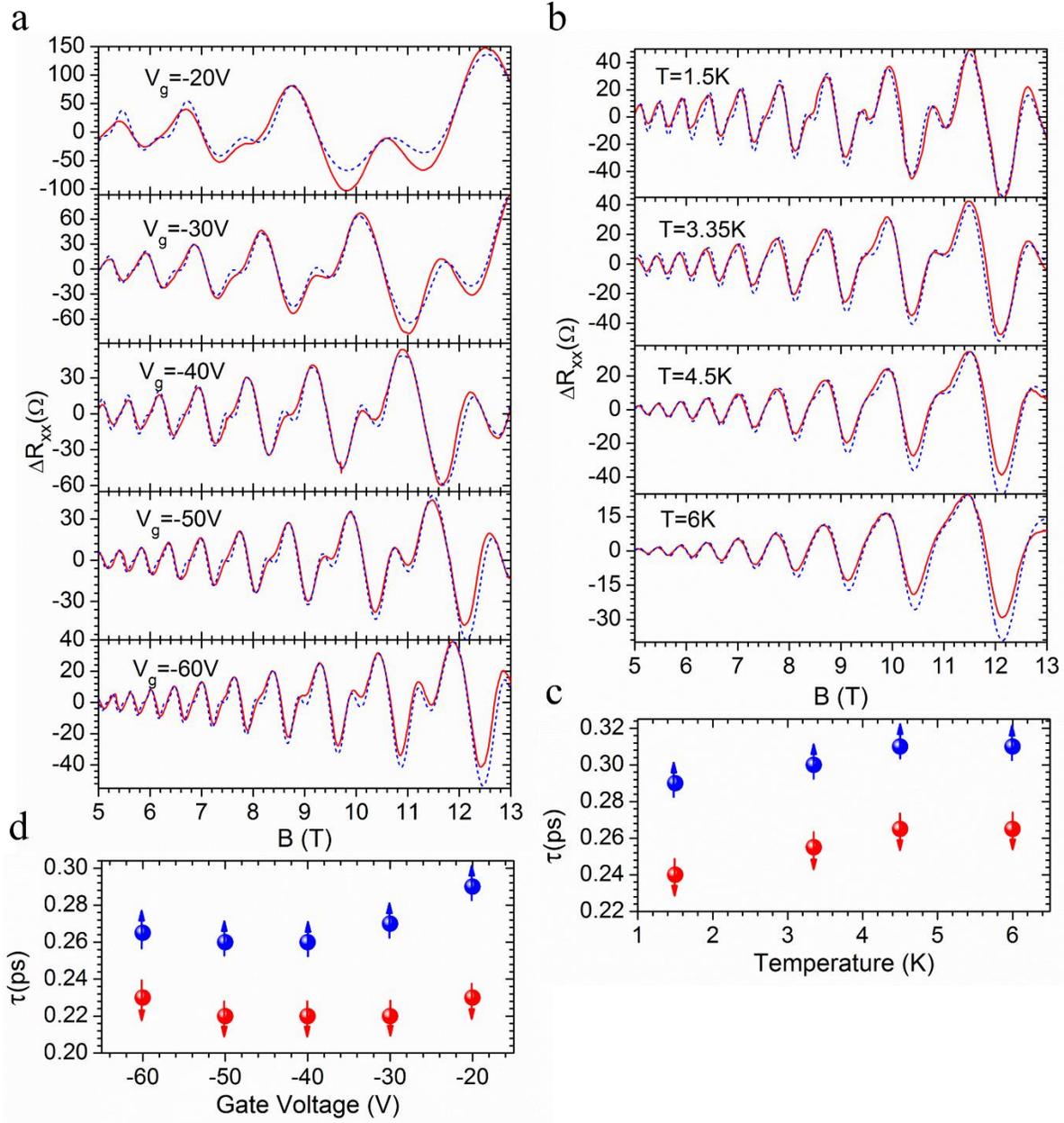

**Fig.4 Fitting results of SdH oscillations according to the spin-resolved LK formalism. (a), (b)** Magnetoresistance (red solid lines) at different gate voltages and temperatures respectively. Blue dashed lines represent the fitting results according to Eq. (1) using m*=0.26 m$_0$; g=3.4 and r =1, 2, 3…20. **(c), (d)** show the spin resolved quantum scattering time extracted from the fitting

results plotted as a function of temperatures and gate voltages respectively. Blue solid dots for spin up and red solid dots for spin down.

**Table.1. Effective mass measured at different carrier densities in BP 2DHG**

| Sample No. | Current direction | Carrier density /$10^{12}$cm$^{-2}$ | Effective mass /$m_0$ |
|---|---|---|---|
| A | X | 4.7 | 0.259 |
|   |   | 2.8 | 0.264 |
| B | X | 4.5 | 0.262 |
|   |   | 2.5 | 0.260 |
| C | Y | 4.7 | 0.260 |
|   |   | 2.9 | 0.257 |
| D | Y | 4.6 | 0.258 |
|   |   | 2.4 | 0.263 |

# Supplementary Materials for

## Quantum Hall Effect in Ultrahigh Mobility Two-dimensional Hole Gas of Black Phosphorus

G. Long, D. Maryenko, J. Y. Shen, S. G. Xu, J. Q. Hou, Z. F. Wu, W. K. Wong, T. Y. Han, J. X. Z. Lin, Y. Cai, R. Lortz, N. Wang

correspondence to: phwang@ust.hk

**Contents**

1. Device fabrication and measurement

2. I-V characteristics of BP devices

3. Measurement of $R_{xx}$

4. Gate voltage dependence of QHE and SdH oscillations

5. Thermal effects on $R_{xx}$

6. Background of SdH oscillations

7. Crystallographic directions of BP channels

8. Effective mass of BP 2DHG

9. The edge-bulk channel coupling

1. **Device fabrication and measurement**

BP flakes are first isolated on heavily doped silicon substrates covered with 300 nm-thick $SiO_2$ in a glove box filled with highly pure nitrogen. The bottom and top h-BN flakes are isolated on silicon substrates and PMMA films, respectively. Then the vacuum in the glove box is kept at $10^{-3}$ Torr. The assembly of the h-BN/BP/h-BN heterostructures is carried out using the manipulators controlled by the panel outside the glove box. Annealing treatment at 550~650K for 15 hrs is necessary to stabilize the heterostructures. Since the heterostructures are formed under vacuum conditions, the charge impurities on the interface between BP and h-BN are further reduced, resulting in further improvement of the device quality. Standard electron beam lithography technique is used to pattern the contact areas the Hall bar devices, followed by a selective plasma etching process to realize ohmic contacts. Electrical measurements are performed using standard lock-in techniques in a cryostat (1.5–300 K with magnetic fields up to 14 T).

## 2. I-V characteristics of BP devices

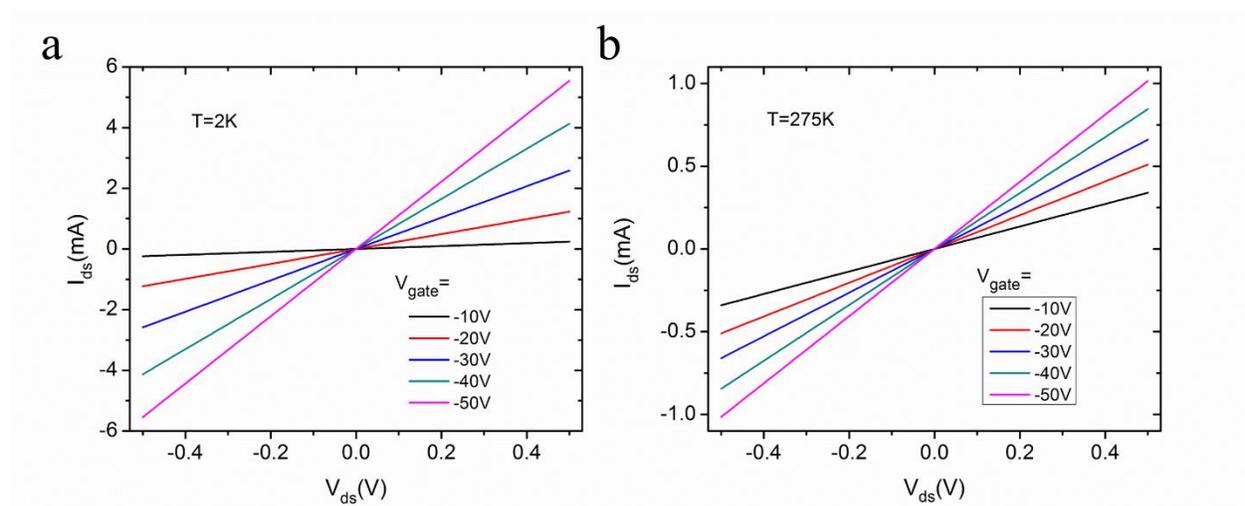

**Fig.S1.** $I_{ds}$-$V_{ds}$ curves obtained by two-terminal configuration of a BP device at 2K (a) and 275K (b) with different gate voltages (Black: -10V, Red: -20V, Blue: -30V, Green: -40V, Pink: -50V). Linear $I_{ds}$-$V_{ds}$ characteristics are detected at both room and cryogenic temperatures, demonstrating that a high-quality of Ohmic contacts between the metal electrodes and BP flakes have been achieved.

## 3. Measurement of Rxx

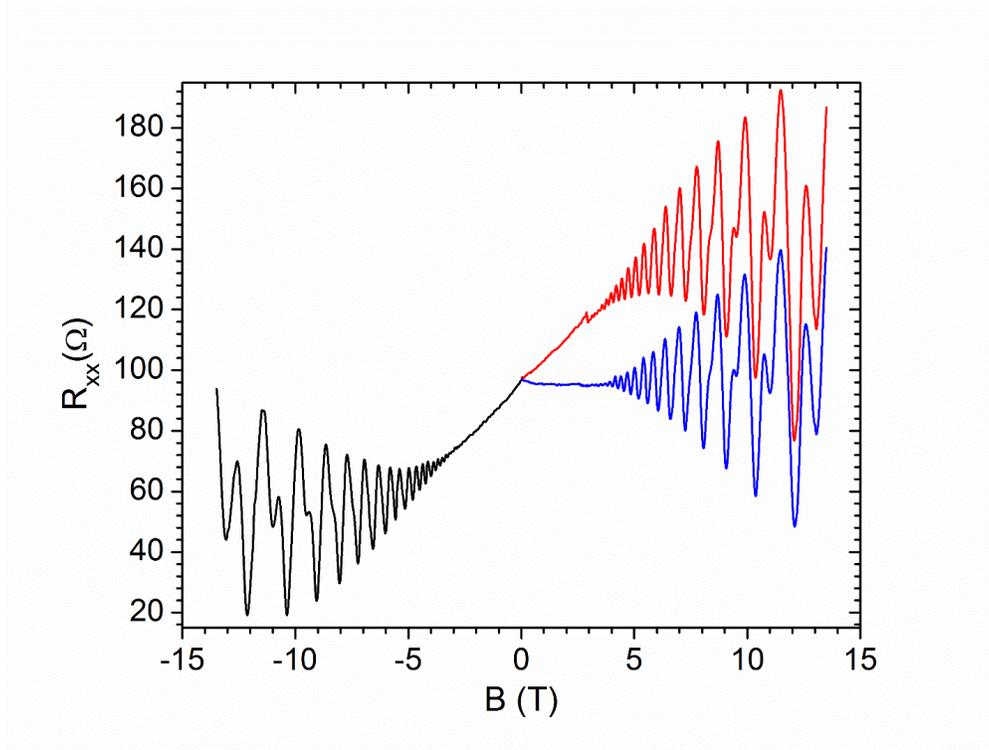

**Fig. S2.** $R_{xx}$ measured under magnetic fields with opposite directions (Black: negative magnetic field; Red: positive magnetic field; Blue: Average data).

Due to the limitation of device fabrication technique, the measured $R_{xx}$ and $R_{xy}$ always mix with each other. $R_{xx}$ and $R_{xy}$ have different parities if magnetic field changes direction, which means:

$$R_{xx}(B) = R_{xx}(-B)$$

$$R_{xy}(B) = -R_{xy}(-B)$$

The mixing component is removed by measuring $R_{xx}$ under magnetic fields with opposite directions.

## 4. Gate voltage dependence of QHE and SdH Oscillatiosns

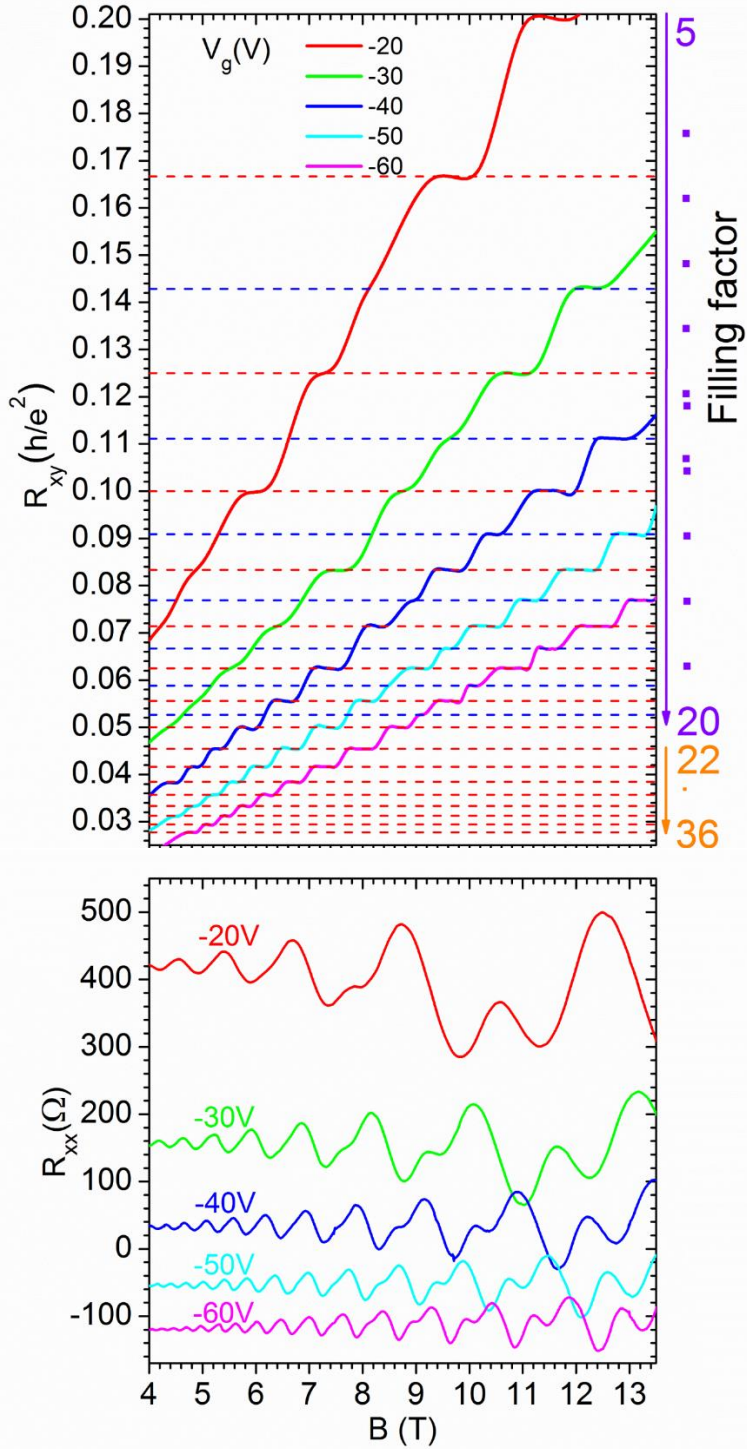

**Fig.S3** Quantum Hall effect measured under different gate voltages. The Hall resistance (top panel) and longitudinal resistance (bottom panel) are plotted as a function of magnetic field at different gate voltages. The measurements are performed at T=1.7K. The red and blue dashed lines in the top panel show the positions of even and odd filling factors respectively. Magnetoresistance curves are shifted vertically by multiples of -50 Ω for clarity.

## 5. Thermal effects on Rxx

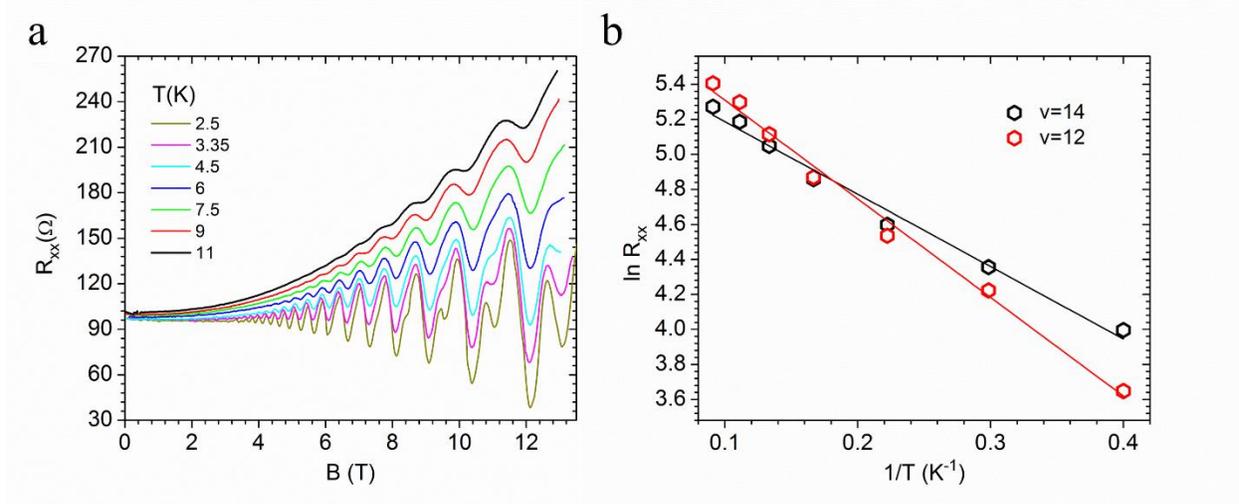

**Fig.S4.** Thermally activated magnetoresistance. **(a)** Magnetoresistance measured at different temperatures showing the features of thermally activated $R_{xx}$. **(b)** $\ln(R_{xx})$ linearly depend on 1/T for filling factors v=12 and v=14.

Fig.S4a shows the detailed features of $R_{xx}$ at different temperatures. The minima of thermally activated $R_{xx}$ can be expressed as $R_{xx}^{\min} \sim \exp(-\Delta E / 2k_B T)$ for even filling factors, where $\Delta E$ denotes the activation energy. In fig.S4b, $\ln(R_{xx}^{\min})$ is plotted as a function of 1/T for filling factors v=12 and v=14. The linear dependence of $\ln(R_{xx}^{\min})$ on 1/T demonstrates a thermally activated effect for $R_{xx}^{\min}$ in our devices. The activation energies are 0.97meV and 0.71meV for v=12 and 14, respectively.

## 6. Background of SdH Oscillations

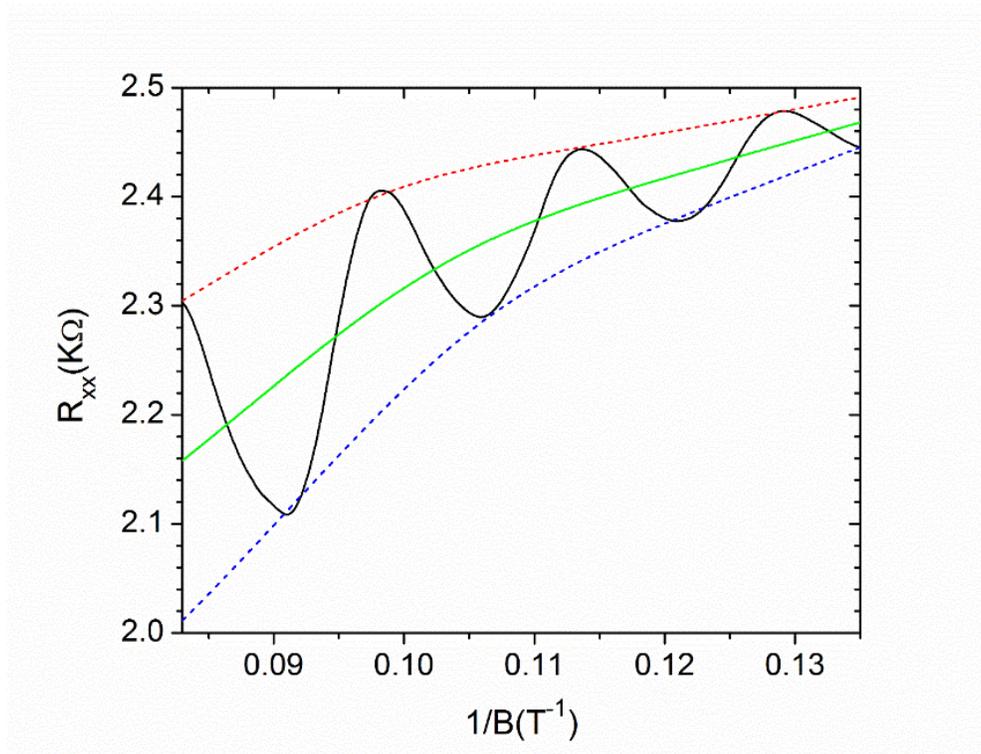

**Fig. S5.** Background of SdH Oscillations. The black solid line displays the measured MR curve of a typical BP device. The red and blue dashed lines represent the up and bottom envelope lines of the MR curve respectively. The average line (the green solid line) of these two curves are taken as the background of the MR curve, which means we substrate the green solid line from measured MR curve to extract the oscillation component before calculating the effective mass.

## 7. Crystallographic directions of BP channels

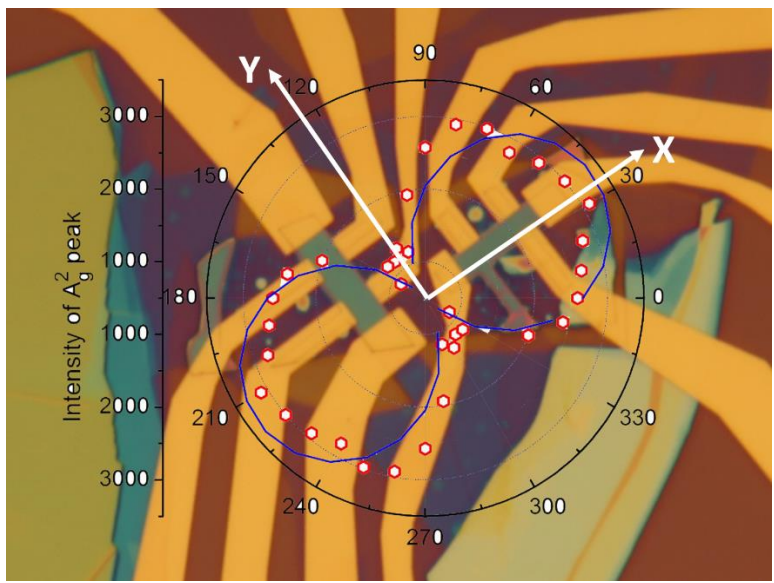

**Fig.S6.** Determination of the crystallographic directions of BP channels by angle-resolved Raman spectroscopy. The polar diagram shows the intensity of $A_g^2$ peak when the polarization direction of the incident laser is in different angles (red dots). The blue line represents the simulation results. The $A_g^2$ peak intensity of BP becomes maximal (minimal) when the laser polarization is in the X (Y) direction of BP (*1*). The crystallographic direction is determined according to the variation trend of the intensity of the $A_g^2$ peak. The crystallographic directions of our samples are indicated by the two white arrows labeled with X and Y.

## 8. Effective mass of BP 2DHG

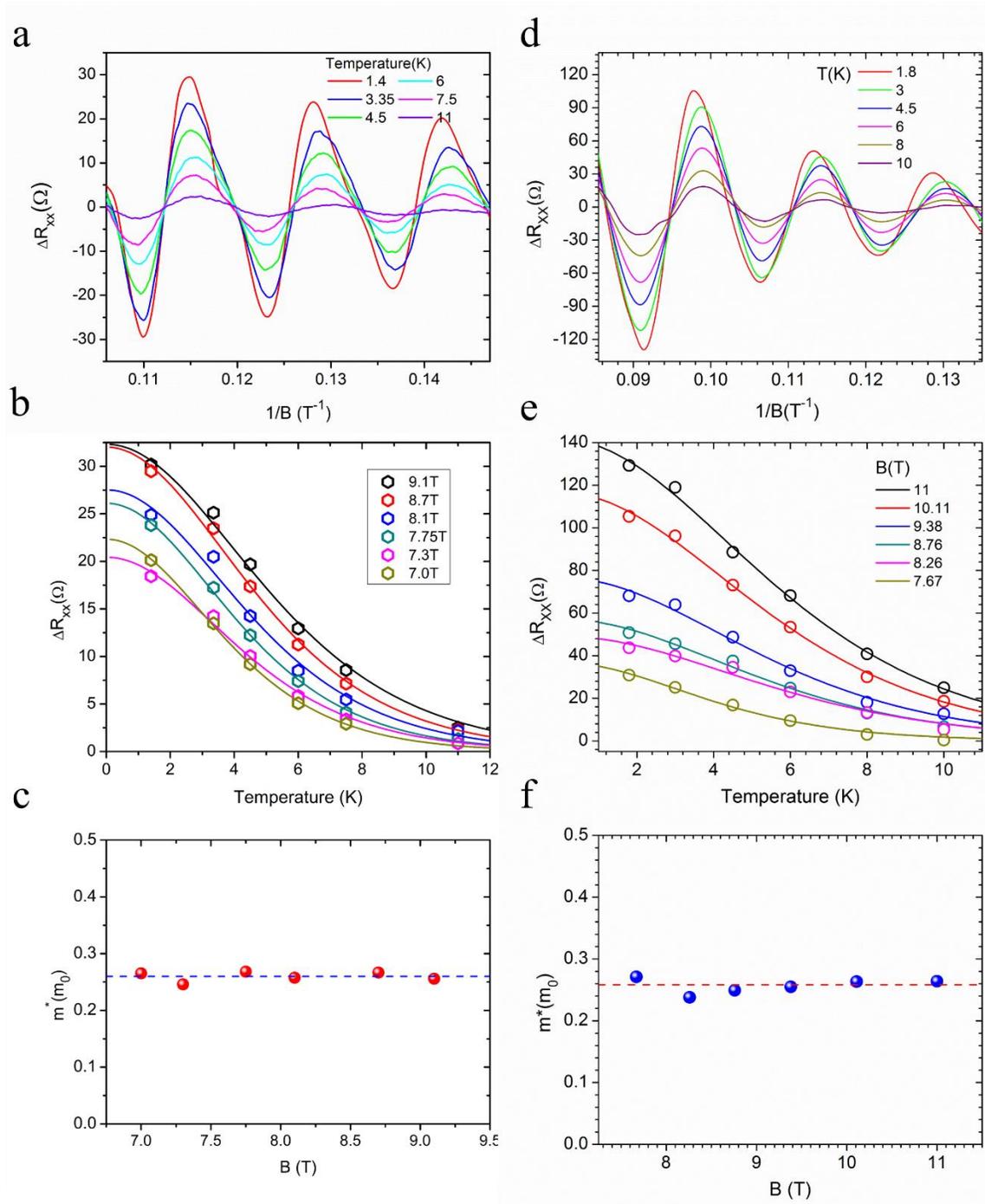

**Fig. S7.** Shubnikov-de Hass oscillations and determination of the effective mass in BP 2DHG. (a), (d) 1/B dependence of SdH oscillation amplitudes ($\Delta R_{xx}$) measured along X and Y

directions at different temperatures. (b), (e) Temperature dependence of SdH oscillation amplitudes measured along X and Y directions under different magnetic fields. The solid lines are fitting results using the LK formula $\Delta R_T \propto \lambda(T)/\sinh(\lambda(T))$ where $\lambda(T) = 2\pi^2 k_B T m^*/\hbar eB$. (c), (f) The cyclotron masses of BP 2DHG obtained from the fitting results of SdH oscillation amplitudes measured from X and Y directions. The dashed lines show the average cyclotron mass.

**Table.S1. Reported effective mass of Black Phosphorus 2DHG**

| Hole effective mass /$m_0$ | Magnetic field /T | Carrier density /$10^{12}$cm$^{-2}$ | Source |
|---|---|---|---|
| 0.27 | 6.5-8.0 | 4 | *Ref. (2)* |
| 0.26-0.31 | 12 | 2.5-5.1* | *Ref. (3)* |
| 0.24±0.02 | 17 | 3.3-5.1 | *Ref. (4)* |
| 0.34±0.02 | 14.3-22.2 | 6.5 | *Ref. (5)* |
| 0.257-0.264 | 7-11 | 2.5-4.7 | *This work* |

* The carrier density was estimated from the frequency of SdH oscillations provided in Ref. (*3*).

## 9. The edge-bulk channel coupling

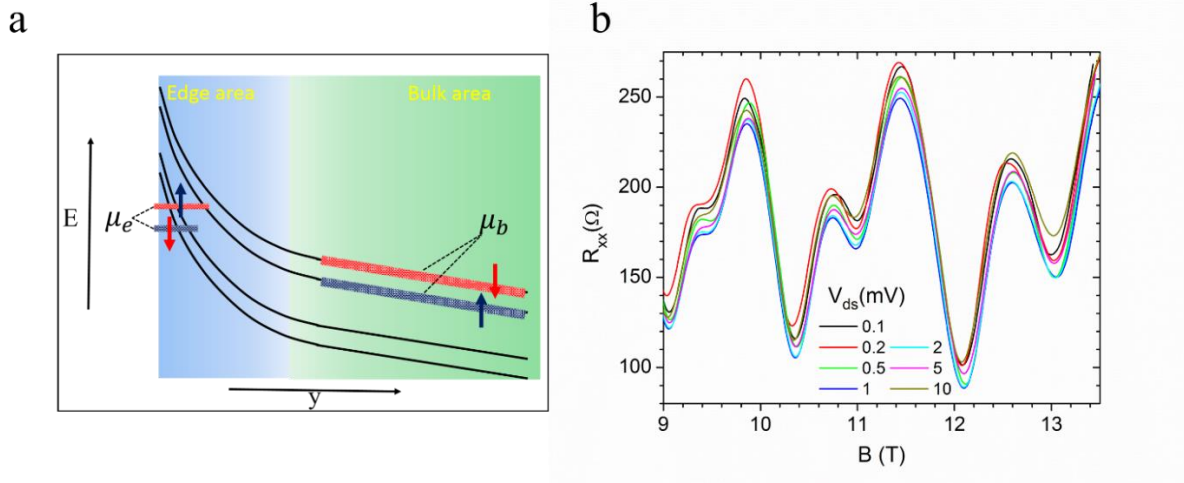

**Fig. S8** Edge-Bulk channel coupling. **(a)** The schematic of the energy level in the vicinity of the sample edge. $\mu_e$ and $\mu_b$ represent the chemical potentials of edge and bulk channels, respectively. The red (blue) arrow indicates the spin polarization of charge carriers. The red (blue) rectangle stands for the chemical potentials of edge and bulk channels, when $E_F$ locates at the centers of spin down (up) landau levels. The spin polarizations of edge and bulk channels are opposite. **(b)** Dependence of SdH oscillations on the excitations ($V_{ds}$) at T=2K.

The spin-resolved transport characteristics measured with $E_F$ fixed at the center of a certain Landau level in two dimensional hole gas systems can indicate the current-induced coupling of the edge and bulk channels (*6-9*). When the Fermi energy ($E_F$) approaches the center of $i^{th}$ Landau level, the $i^{th}$ channel carriers produce a current in the bulk (extended states), which gives rise to the finite resistivity along the sample (*9*). According to Ref.(*9*), the finite resistivity of sample channels may involve charge carrier transitions from the edge channels into the bulk followed by the charge carriers transitions through bulk channels to the opposite sample edge. Due to the different current carrying abilities between edge and bulk states, a chemical potential difference is induced between the edge channel and the adjacent part of the bulk by the applied current (*9*) (Fig.S8a). The increased energy of edge states may cause the carriers transitions between edge states and bulk states. Different behaviors of the carriers with different spin orientations have been ascribed to their different transition rates between edge and bulk channels. The transition rates are determined by the spatial separations ($\Delta l$) between the upmost edge channel and the bulk channel. If the quantum levels at $E_F$ are occupied by spin up (down) charge carriers, $\Delta l$ scales with $\hbar\omega$ ($g\mu_B B$). So the difference between $\hbar\omega$ and $g\mu_B B$ leads to the different behaviors of charge carriers with different spin orientations. According to P. Svoboda, *et al.*, the spin-orbit coupling results in mixing states between states with different spin orientations. Then, the transitions between edge states and bulk states become possible (*9*). However, in our case, the spin-orbit coupling is too weak to generate the mixing states (*10, 11*). The transitions between spin polarized states are therefore forbidden. Experimentally, the most remarkable features of the coupling for the edge and bulk channels – the spin-dependent nonlinearity of peak values with respect to the excitation voltage ($V_{ds}$) has not been observed (Fig.S8b). Fig. S8b demonstrates that the SdH oscillations observed in our samples are independent of different excitations ($V_{ds}$). Clearly, the edge-bulk coupling (if any) has no impact on our few-layer BP devices.

**Supplementary References:**